\documentclass[aps,prb,twocolumn,a4paper,10pt,superscriptaddress,longbibliography]{revtex4-1}

\usepackage{float}
\usepackage{graphicx}
\usepackage{graphics}
\usepackage{amsmath}
\usepackage{amssymb}
\usepackage{amsfonts}
\usepackage{dsfont}
\usepackage{braket}
\usepackage{color}
\usepackage{braket,slashed}
\usepackage[mathscr]{euscript}
\usepackage{hyperref}
\usepackage{epstopdf}
\usepackage{subfigure}
\usepackage{xfrac}
\usepackage{bm}
\usepackage{natbib}
\usepackage{kantlipsum}

\allowdisplaybreaks[1]

\renewcommand{\d}{\ensuremath{\mathrm{d}}}
\newcommand{\e}{\ensuremath{\mathrm{e}}}

\begin{document}

\title{Quasi-particle interactions in frustrated Heisenberg chains}

\author{Laurens Vanderstraeten}
\affiliation{Ghent University, Department of Physics and Astronomy, Krijgslaan 281-S9, B-9000 Gent, Belgium}
\author{Jutho Haegeman}
\affiliation{Ghent University, Department of Physics and Astronomy, Krijgslaan 281-S9, B-9000 Gent, Belgium}
\author{Frank Verstraete}
\affiliation{Ghent University, Department of Physics and Astronomy, Krijgslaan 281-S9, B-9000 Gent, Belgium}
\affiliation{Vienna Center for Quantum Science, Universit\"at Wien, Boltzmanngasse 5, A-1090 Wien, Austria}
\author{Didier Poilblanc}
\affiliation{Laboratoire de Physique Th\'{e}orique, IRSAMC, CNRS and Universit\'{e} de Toulouse, UPS, F-31062 Toulouse, France}

\begin{abstract}
Interactions between elementary excitations in quasi-one dimensional antiferromagnets are of experimental relevance and their quantitative theoretical treatment has been a theoretical challenge for many years. Using matrix product states, one can explicitly determine the wavefunctions of the one- and two-particle excitations, and, consequently, the contributions to dynamical correlations. We apply this framework to the (non integrable) frustrated dimerized spin-1/2 chain, a model for generic spin-Peierls systems, where low-energy quasi-particle excitations are bound states of topological solitons. The spin structure factor involving two quasi-particle scattering states is obtained in the thermodynamic limit with full momentum and frequency resolution. This allows very subtle features in the two-particle spectral function to be revealed which, we argue, could be seen, e.g., in inelastic neutron scattering of spin-Peierls compounds under a change of the external pressure.
\end{abstract}

\maketitle

\section{Introduction}

The physics of one-dimensional quantum magnets has been the subject of extensive experimental and theoretical study. Because of the reduced dimensionality, the quantum fluctuations in these spin chains are especially strong, with a rich variety of quantum phases as a result. The quasi-particles in these systems are collective excitations \cite{Giamarchi2004}, in no way connected to some free particle limit, and exhibit exotic physical properties such as fractional quantum numbers, non-trivial scattering properties, soliton confinement, bound state formation, etc.
\par In recent years, the interactions between these quasi-particles have come within the scope of experiments.  Most importantly, the spectral functions as measured in inelastic neutron scattering (INS) experiments have important contributions from many-particle states \cite{Ain1997, Notbohm2007, Lake2013, Stone2014a}. Also, the thermal broadening of one-particle signals can only be accounted for by many-particle processes \cite{Essler2008, Tennant2012}. Thirdly, the critical properties of a magnetized spin chain or ladder is understood by identifying it as a condensed gas of interacting magnons on a strongly-correlated background state \cite{Klanjsek2008, Ruegg2008, Schmidiger2012, Giamarchi2008}. Lastly, the properties of quasi-particles can be probed in cold-atom experiments \cite{Cheneau2012, Fukuhara2013a, Jurcevic2014} and their interactions are important \cite{Fukuhara2013b, Jurcevic2015}.
\par These experimental advances have spurred the development of a number of theoretical tools for simulating the low-energy dynamics of one-dimensional spin systems. Exact diagonalization is the most straightforward and unbiased approach \cite{Hayward1996, Poilblanc1997, Augier1997, Laflorencie2004, Capponi2007}, but, due to its exponential scaling, limited to small system sizes. When frustration is present in the system, quantum Monte Carlo methods are plagued by the sign problem. In recent years, the scope of the density-matrix renormalization group (DMRG) \cite{White1992} has been extended to study time evolution and, through a Fourier transform, spectral functions \cite{White2008a}. This method has the advantage of being completely generic and has an efficient scaling in system size, but is limited in its resolution; it is the growth in entanglement during real-time evolution that limits the resolution of the spectral functions. Other DMRG approaches, based on Lanczos-methods \cite{Dargel2012} or Chebyshev expansions \cite{Holzner2011}, have similar defects. Yet, if one is only interested in low-energy dynamics, this entanglement growth seems counter-intuitive as all states in this low-energy sector are characterized by a small entanglement entropy \cite{Hastings2006, Masanes2009, Eisert2010a}.
\par A different strategy consists in targeting the low-lying particle excitations explicitly. For integrable systems, all excited states can be constructed exactly and it has been shown how to compute their spectral weights in an efficient way \cite{Caux2005, Caux2005a, Caux2009}. Alternatively, if the system can be connected perturbatively to a trivial non-interacting limit, perturbative continuous unitary transformations provide access to the non-trivial properties of low-lying excitation spectra \cite{Trebst2000, Knetter2001, Schmidt2003, Schmidt2004}. Both approaches have the advantage that they have a built-in particle picture, but, for these particles to be well-defined, they need an extensive number of conserved quantities or a well-defined non-interacting limit to perturb from.
\par It can be shown in full generality, however, that a particle picture should hold for any gapped quantum lattice system \cite{Haegeman2013a}. It is by exploiting the formalism of matrix product states (MPS)  \cite{Verstraete2008a, Schollwock2011a}, that an accurate variational ansatz for particle-like excitations in one-dimensional spin systems was first constructed \cite{Haegeman2012a, Haegeman2013b}. Within this approach, excitations are seen as momentum superpositions of local perturbations on an MPS ground state---it can be seen as an extension of the single-mode approximation \cite{Bijl1941, Feynman1953a, *Feynman1954, Girvin1985, *Girvin1986, Arovas1988}---and can be naturally interpreted as effective particles living on a strongly-correlated vacuum state. Subsequently, it has been shown how to model the interactions between these particles and how to compute the two-particle S matrix \cite{Vanderstraeten2014, Vanderstraeten2015a}. This information on the two-particle interactions allowed for an effective many-particle description of e.g. magnon condensation under strong magnetic fields.
\par With this approach, one can explicitly determine the wavefunctions of the one- and two-particle excitations, and, consequently, the contributions to dynamical correlations \cite{Vanderstraeten2015a}. In this way, it allows for an efficient evaluation of spectral functions by combining the advantages of the aforementioned methods. Indeed, because it targets the exact particle excitations in the thermodynamic limit, this framework has access to the low-lying part of the spectral function with full resolution in momentum and frequency. Moreover, as it is based on matrix product states, it is applicable to generic quantum spin chains with relatively low numerical resources.
\par In Ref.~\onlinecite{Vanderstraeten2015a} a detailed account of this method has been presented. In the present contribution, we show for the first time its versatility by applying the method to the frustrated spin-1/2 Heisenberg chain. We track the effect of the quasiparticle interactions on the spectral function in the two-particle band. Before that, we give some non-trivial benchmark results on the spectral function of the gapped XXZ antiferromagnet, which can be directly compared with Bethe ansatz results.

\section{Benchmark: the XXZ chain}

\begin{figure} \centering
\includegraphics[width=\columnwidth]{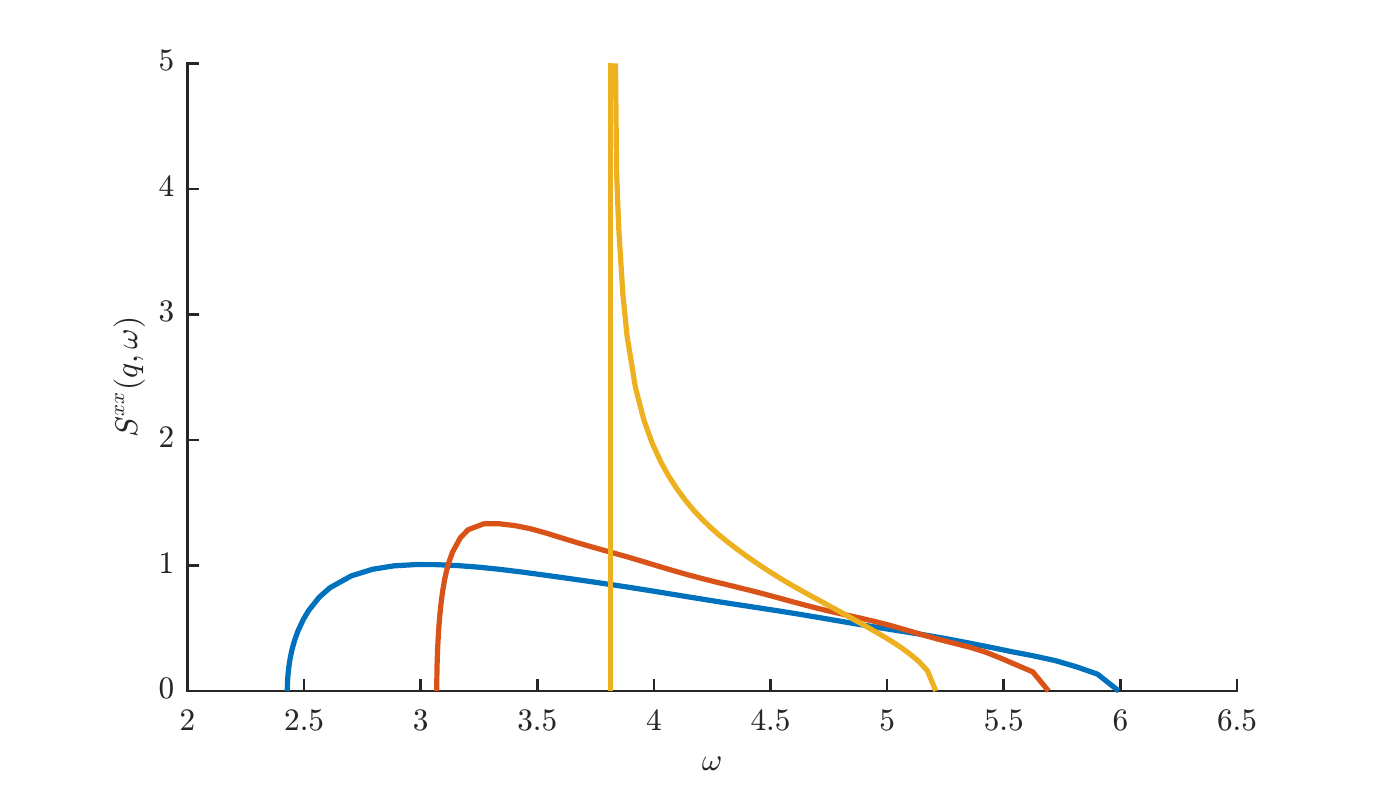}
\caption{The spectral function $S^{xx}(q,\omega)$ of the XXZ chain with $J_z=4$; momentum cuts at $q=\pi/8$ (blue),  $q=2\pi/8$ (red) and  $q=3\pi/8$ (orange). The orange curve shows a divergence at the band's edge. Note that the plotting range cuts off the orange curve; because our method works directly in the thermodynamic limit, we can reproduce this divergence to arbitrary precision. A comparison with Fig. 5 in Ref. \onlinecite{Caux2008} shows a very strong similarity.}
\label{fig:xxz}
\end{figure}

The XXZ spin-1/2 Heisenberg chain is defined by the Hamiltonian
\begin{equation*}
H = \sum_n \left( S_n^x S_{n+1}^x + S_n^y S_{n+1}^y + J_z S_n^z S_{n+1}^z \right).
\end{equation*}
and is gapped for $J_z>1$. The model is integrable \cite{Gaudin2014} and the spectrum can be computed exactly with the Bethe ansatz. In this regime there is a two-fold degenerate ground state with a finite N\'{e}el order parameter, and the elementary excitations are spinons carrying a fractional spin of 1/2 \cite{Faddeev1981}. The first contributions to the spectral function
\begin{equation*}
S^{xx}(q,\omega) = \int\d t \sum_n \e^{i\omega t}\e^{iqn} \langle S_n^x(t) S_0^x(0) \rangle
\end{equation*}
come from two-spinon states, which can be computed exactly \cite{Bougourzi1998,Caux2008} using Bethe-ansatz techniques.
\par In Ref.~\onlinecite{Haegeman2012a} the dispersion relation of the individual spinon excitations were computed to very high precision. With our method, two-spinon scattering states can be determined and their contributions to the spectral function evaluated. In Fig.~\ref{fig:xxz} we have plotted three momentum cuts of $S^{xx}(q,\omega)$ for $J_z=4$, which can be compared to the Bethe ansatz results from Ref.~\onlinecite{Caux2008}. The correspondence is very good, which confirms the accuracy of our method.

\section{The frustrated Heisenberg chain}

\begin{figure}
\includegraphics[width=0.7\columnwidth]{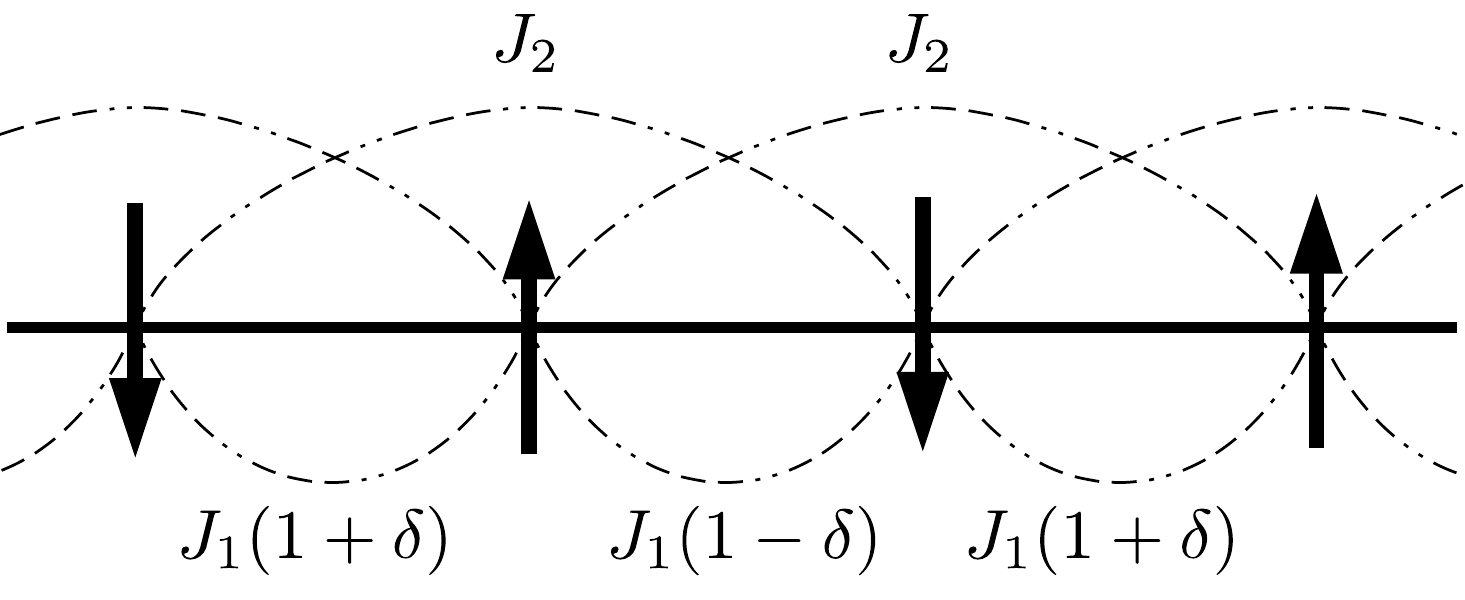}
\caption{The frustrated and dimerized chain.}
\label{fig:chain}
\end{figure}
\begin{figure}
\includegraphics[width=0.7\columnwidth]{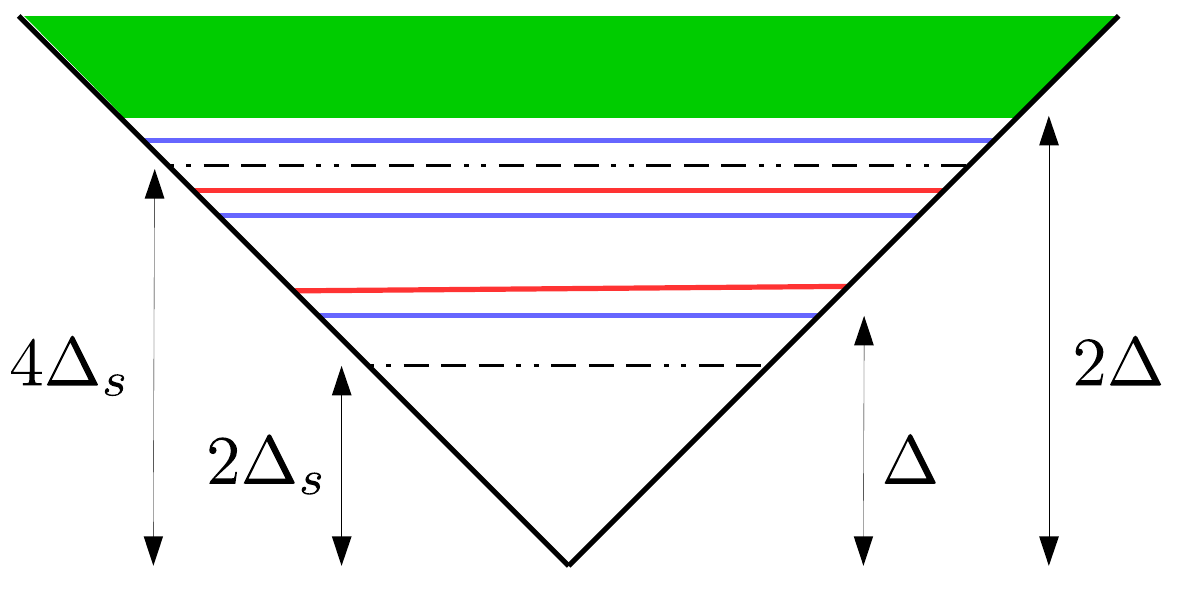}
\caption{A sketch of the bound-state spectrum; different excitations can be seen as bound state solutions of a linear potential. The strength of the potential ($\propto\delta$) can be tuned more or less independently from the mass $\Delta_s$ of the confined solitons. A number of triplet (blue) and singlet (red) $s\bar{s}$ bound states appear above the two-soliton energy $2\Delta_s$; these appear at energies $2\Delta_s+E_{b,i}$, with $E_{b,i}$ the (positive) binding energy of the $i$'th solution of the linear confining $s\bar{s}$ potential. Above the edge of the $s\bar{s}$-$s\bar{s}$ continuum (green) these solutions are no longer stable against decay into scattering states of two $s\bar{s}$ states. Note that the stability of a bound state depends on its momentum (see Fig.~\ref{fig:spectrum}).}
\label{fig:schematic}
\end{figure}

\begin{figure*} \centering
\subfigure[$\;\delta=0.05$]{\includegraphics[width=\columnwidth]{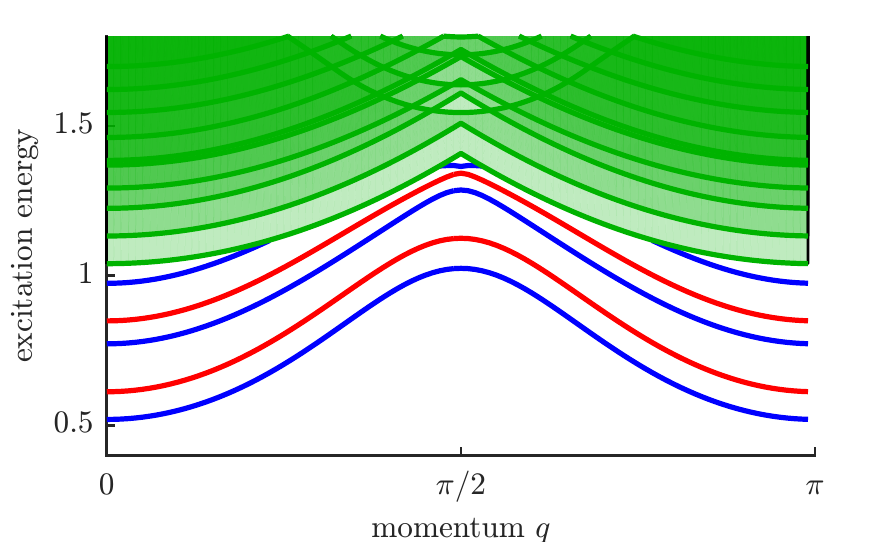}\label{a}}
\subfigure[$\;\delta=0.08$]{\includegraphics[width=\columnwidth]{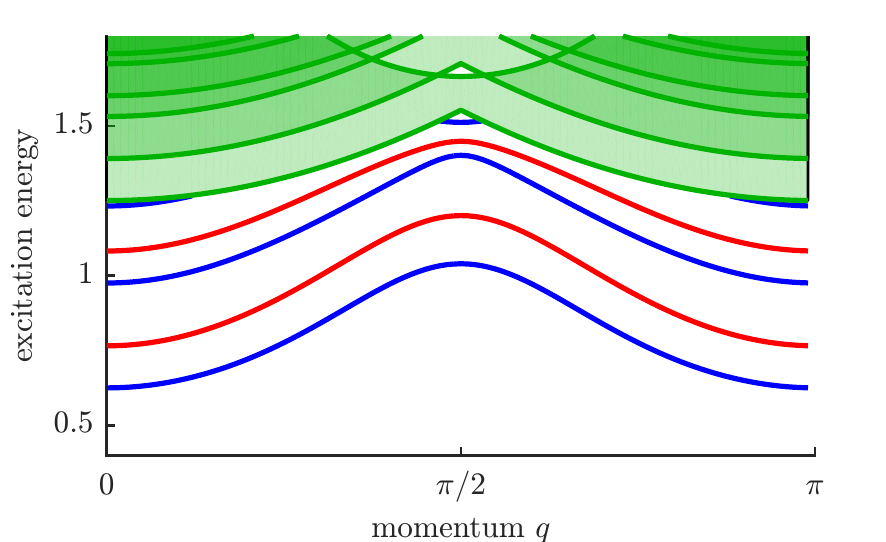}\label{b}}\\
\subfigure[$\;\delta=0.11$]{\includegraphics[width=\columnwidth]{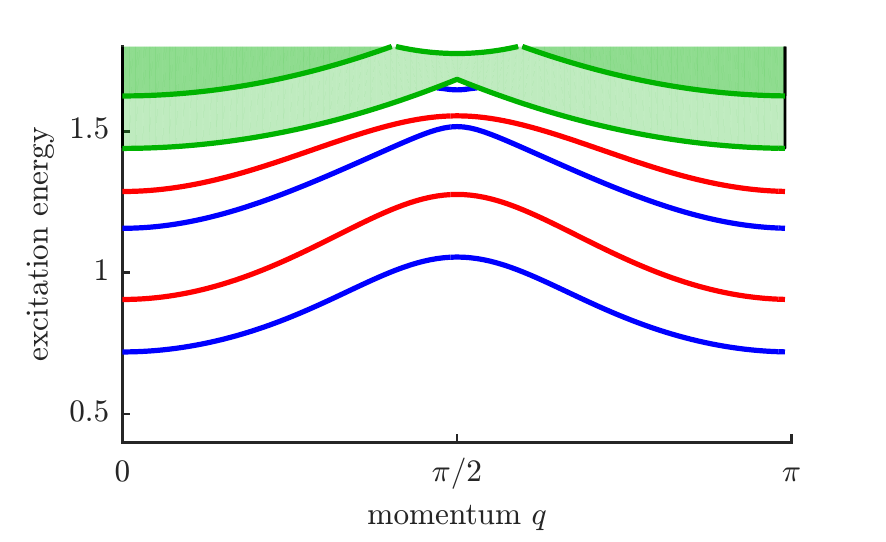}\label{c}}
\subfigure[$\;\delta=0.13$]{\includegraphics[width=\columnwidth]{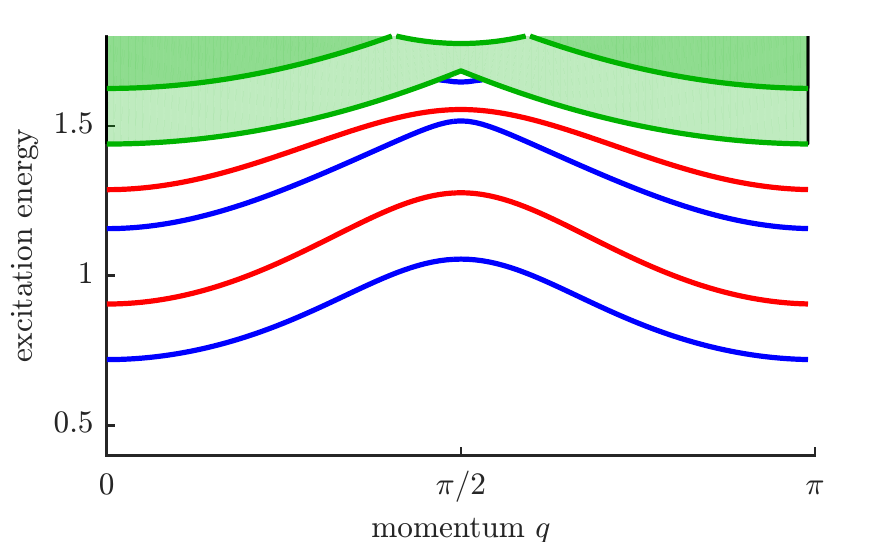}\label{d}}
\caption{The elementary excitation spectrum for $J_1=1$, $J_2=1/2$ and four different values of $\delta$. Because the ground state has a two-site unit cell, the Brillouin zone is confined to momenta $q\in[0,\pi)$ (in units of inverse lattice spacing $a^{-1}$). For all values of $\delta$ one can observe two stable triplet (blue) and singlet (red) $s\bar{s}$ bound states over the full momentum range. For small values of $\delta$ a third triplet is stable at the minimum. At the center of the Brillouin zone a third triplet state is stable for all values of $\delta$.}
\label{fig:spectrum}
\end{figure*}

An excellent model to study the signature of quasi-particle interactions in the spectral function is the frustrated and dimerized spin-1/2 Heisenberg chain, given by the Hamiltonian (see Fig.~\ref{fig:chain})
\begin{equation*}
H =  J_1\sum_n(1 + \delta(-1)^n) \vec{S}_n \cdot \vec{S}_{n+1} + J_2 \sum_n \vec{S}_n \cdot \vec{S}_{n+2} .
\end{equation*}
It has been shown \cite{Augier1998a} that spin-Peierls compounds such as $\text{CuGe}_2$ and $\text{NaV}_2\text{O}_5$ can be described with this model, where the dimerization term arises from a three-dimensional coupling of the spin chains. The physics of spinon confinement due to this three-dimensional coupling has recently attracted a lot of experimental and theoretical attention \cite{Lake2009, Coldea2010, Wang2015}.

\par Without explicit dimerization ($\delta=0$) it is known \cite{Haldane1982, Haldane1982a} that a gap opens at $J_2/J_1\approx0.2412$ \cite{Okamoto1992}, accompanied by a spontaneous lattice dimerization and a twofold-degenerate ground state. At $J_2/J_1=1/2$ the Majumdar-Ghosh model \cite{Majumdar1969a, *Majumdar1969b, *Majumdar1970} is retrieved, for which the ground state is an exact product state of dimers. Throughout the dimerized phase, the elementary excitations can be pictured as dressed defects in the dimerization pattern \cite{Shastry1981, Caspers1984} and behave as topological solitons $s$ and anti-solitons $\bar{s}$ interpolating between the two ground states. The solitons have spin 1/2. No bound states occur \cite{Sorensen1998}, so the physical spectrum starts with a soliton/anti-soliton scattering continuum at $2\Delta_s$ (with $\Delta_s$ the soliton gap).
\par The dimerization $\delta$ favours one of the two ground states and confines the solitons into bound states with a linear potential between a $s\bar{s}$ pair \cite{Sorensen1998, Affleck1998, Augier1999}. This implies that the $s\bar{s}$ continuum is split up into a stack of discrete triplet and singlet bound states (see Fig.~\ref{fig:schematic}). If $\Delta$ is the energy of the lowest-lying triplet, a two-triplet continuum will start at $2\Delta$, such that $s\bar{s}$ bound states with an energy above this threshold will not be stable. This is the effect of string breaking: if the energy cost of having the wrong ground state between the $s\bar{s}$ is too high, the bound state will decay to a $s\bar{s}$-$s\bar{s}$ pair. Not too far up in the continuum, however, we expect that unstable, yet long-lived, bound states will leave their signature on the spectral function.


\begin{figure} \centering
\includegraphics[width=\columnwidth]{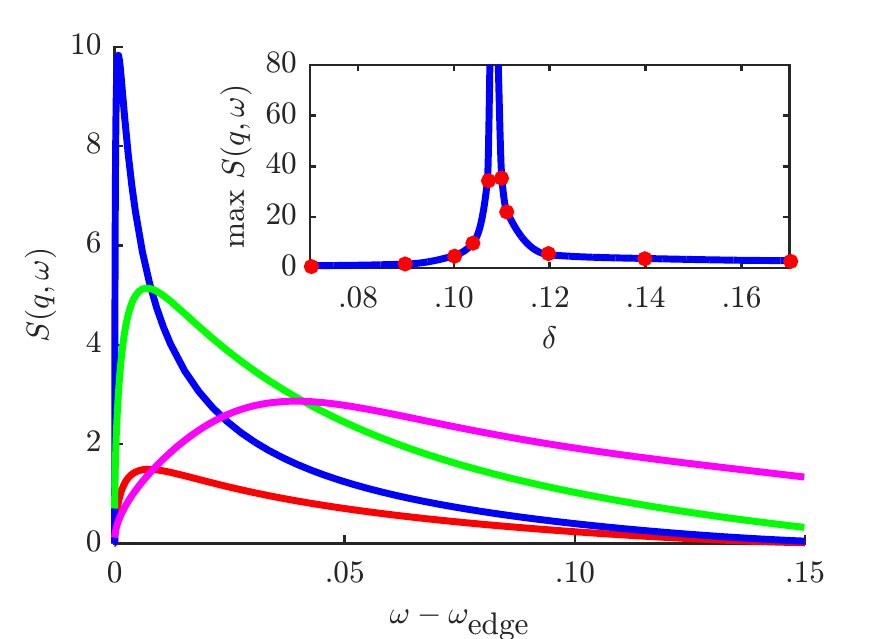}
\caption{The spectral function $S(q,\omega)$ at momentum $q=\pi$ just above the first triplet-triplet threshold for $J_1=1$, $J_2=1/2$, and different values of the dimerization: $\delta=0.09$ (red), $\delta=0.104$ (blue), $\delta=0.12$ (green) and $\delta=0.17$ (magenta); the energies $\omega$ have been shifted by twice the triplet gap $2\Delta(\delta)$. For $\delta=0.09$ the third triplet bound state (at the minimum of the dispersion relation) falls below the continuum, but, as the gap becomes smaller, the peak above the threshold becomes larger. As the bound state enters the continuum, the spectral function diverges. For larger $\delta$ the bound state decays into two $s\bar{s}$ states and becomes a resonance in the spectral function. This resonance travels to higher energies as $\delta$ is further increased. Inset: The maximum of the spectral function (above $\omega_{\rm edge}$) as a function of $\delta$ showing
a divergence at $\delta\approx0.11$ exactly where the third bound state becomes unstable. The blue line is a guide to the eye, the largest data points fall outside the plotting region.}
\label{fig:delta}
\end{figure}

\par We will study this $s\bar{s}$-$s\bar{s}$ continuum through a computation of the spectral function for small dimerization $\delta$ such that the underlying soliton physics can be observed. We will look at the momentum-frequency resolved spin-spin dynamical correlation function as observed in INS, defined as
\begin{equation*}
S(q,\omega) = \int\! \d t \, \e^{i\omega t}\sum_n \e^{iqn} \langle S_n^-(t) S_0^+(0) \rangle
\end{equation*}
where $S_n^{+/-}(t)$ are the ladder operators at site $n$ in the Heisenberg picture and the expectation value $\langle\dots\rangle$ is with respect to the ground state. All results were obtained by applying the MPS framework as developed in Ref. \onlinecite{Vanderstraeten2015a}.
\par In Fig. \ref{fig:spectrum} the elementary excitation spectrum is plotted for $J_2=1/2$ and different values of the dimerization. Whereas two triplet and singlet $s\bar{s}$ bound states are stable over the full Brillouin zone for all values of $\delta$, a third triplet excitation is stable only for small values of $\delta$ and close the dispersion's minimum; it merges into the continuum and loses stability for larger momenta. At the zone center, a triplet emerges again from the two-particle continuum. This bound state was also observed in systems with larger dimerization \cite{Schmidt2004} with the use of perturbative continuous unitary transformations, starting from the isolated dimer limit ($\delta\rightarrow\infty$). Therefore, it seems plausible that the physical origin of this bound state is not connected to the underlying soliton physics (which is only valid for small $\delta$).
\par Let us first focus on the stability of the third triplet bound state at the minimum of the dispersion as a function of $\delta$. In Fig. \ref{fig:delta} we have plotted momentum slices of $S(q,\omega)$ at $q=\pi$ inside the two-particle continuum. For $\delta=0.09$ most of the spectral weight is in the $\delta$-peak of the bound state. As $\delta$ is increased, the bound state comes closer to the continuum at $\omega_{\rm edge}=2\Delta(\delta)$, and the two-particle continuum gains spectral weight and becomes sharply peaked. Just as the bound state enters the continuum, the spectral function diverges; by plotting the maximum of the spectral function as a function of $\delta$ [inset of Fig.~\ref{fig:delta}], the exact value for $\delta$ can be pinpointed for which the third triplet $s\bar{s}$ becomes unstable. For larger $\delta$, the maximum of the spectral function travels through the continuum as a signature of an unstable yet long-lived $s\bar{s}$ state.
\par
\begin{figure*}\centering
\includegraphics[width=1.7\columnwidth]{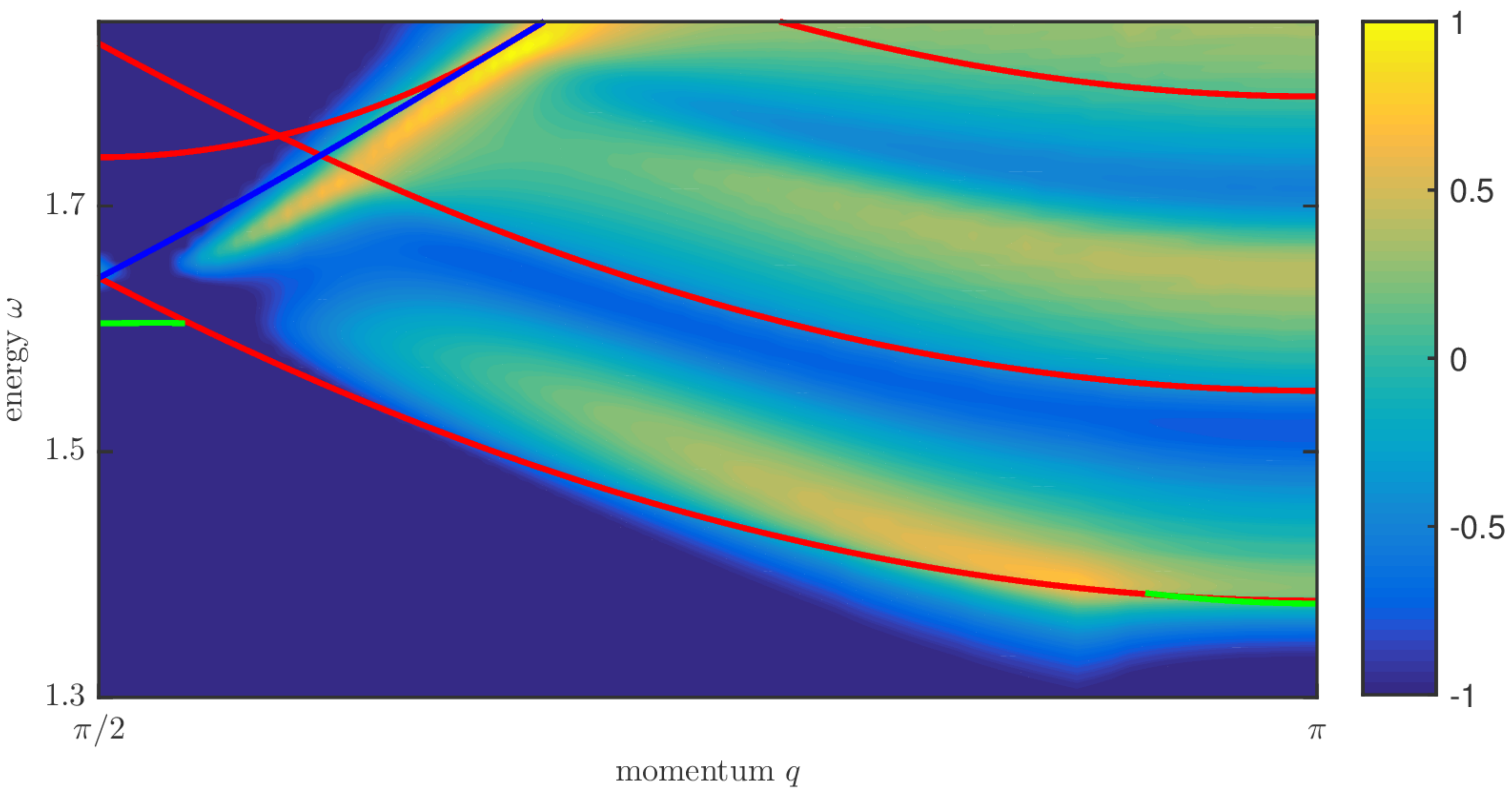}
\caption{The logarithm of the two-particle spectral function $\log_{10}S(q,\omega)$ in the second half of the Brillouin zone for $J_1=1$, $J_2=1/2$ and $\delta=0.1$. Only the two-particle contributions are plotted, the sharp $\delta$-functions of the one-particle states are not shown. A small imaginary frequency $\epsilon=0.01$ is added for aesthetic reasons (note that we have full resolution in $q$ and $\omega$ as our methods work in the thermodynamic limit, see the other plots) and to mimic experimental resolution. We have also plotted the edges of the different two-particle continua in red. The lower edge of the first band consists of states with momenta $\kappa_1=\pi/2+q/2$ and $\kappa_2=-\pi/2+q/2$. The blue line indicates the states with equal individual momenta $\kappa_{1,2}=q/2$, leading to a diverging density of states on that line. Between the blue line and the upper edge of the continuum is the folding region, where different combinations of one-particle states can give rise to the same total momentum and energy. The green line follows the dispersion of the third $s\bar{s}$ bound state where it is stable. We can see that the dispersion of the higher two-particle continua gives a banded structure to the spectral function.}
\label{fig:spectral}
\end{figure*}

 Fig.~\ref{fig:spectrum} illustrated that the bound state loses stability for larger momenta. In Fig.~\ref{fig:spectral} the two-particle spectral function is plotted in the second half of the Brillouin zone (which carries most of its weight) for $J_2/J_1=1/2$ and $\delta=0.1$, for which the third bound state is stable at the minimum of the two-particle continuum. We now observe that, as the momentum moves away from the minimum, the bound state enters the continuum as a sharp resonance and survives throughout a large portion of the Brillouin zone.
\par Further up in frequency, the two-particle spectral function shows a non-trivial structure, because of the different overlapping two-particle continua. In the absence of special symmetries, there are non-zero scattering amplitudes linking the different two-particle sectors. As a result, the eigenstates will mix up these sectors and the spectral function obtains its characteristic banded structure.

\par Another strong resonance appears when the bound state around the zone center enters the continuum. Surprisingly, the bound state does not connect immediately to the resonance at the minimum of the dispersion relation, as another resonance seems to run away with all the spectral weight. This other resonance can be explained as a combined effect from (i) the attractive interaction between the triplet bound states, and (ii) a divergence of the density of states within the continuum. The latter is a consequence of the folding inside the two-triplet continuum: there are regions inside the scattering continuum for which there are two different combinations of one-triplet states that give rise to the same total momentum and energy. The boundary lines of this folding region exhibit, just like the edges of the two-particle continuum, a square-root divergence in the density of states. Fig.~\ref{fig:spectral} shows, however, that the resonance does not coincide with this line. The reason is that the particles have an attractive interaction and, consequently, a negative binding energy which brings the resonance down in energy. This binding energy goes to zero as the resonance travels towards the upper edge of the continuum.

\par
\begin{figure} \centering
\includegraphics[width=\columnwidth]{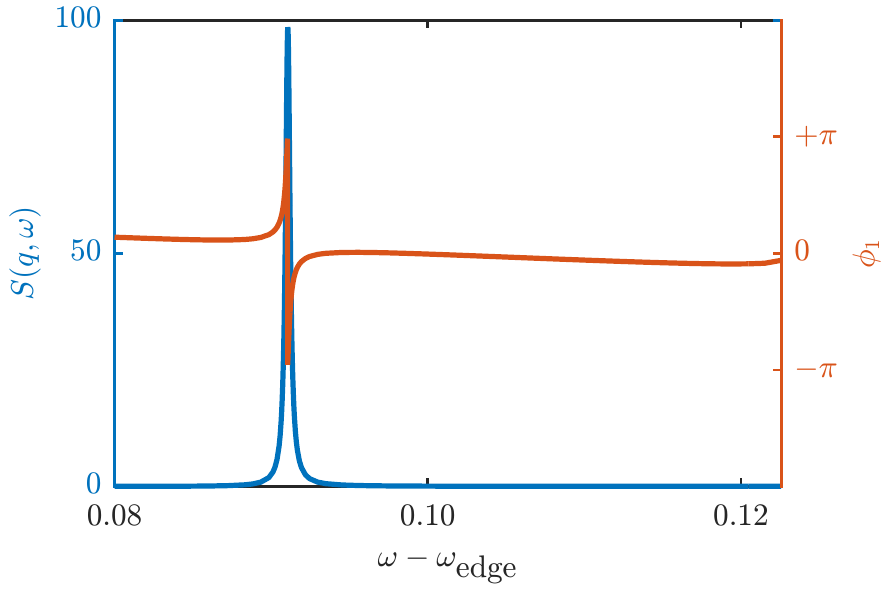}
\caption{The spectral function $S(q,\omega)$ (blue) and the scattering phase in the $S=1$ sector $\phi_1(q,\omega)$ (red) as a function of the rescaled energy $\omega-\omega_\text{edge}$ and for fixed momentum $q=0.56\pi$, where $\omega_\text{edge}$ is the lower edge of the two-particle continuum. We can observe that the scattering phase rotates rapidly at the point where the resonance is located, signalling a strong interaction between the particles at this point. Off the resonance, the dispersion of the scattering phase behaves smoothly. At $\omega-\omega_\text{edge}\approx0.123$ one enters the folding region (with a diverging density of states); beyond that point the S matrix is a $18\times18$ matrix, which can no longer be written in terms of three phases.}
\label{fig:smatrix}
\end{figure}

 In order to corroborate this picture, the particle interactions can be further characterized by studying the two-triplet S matrix. In the case of triplet-triplet scattering, this S matrix is a $9\times9$ unitary matrix, which can be diagonalized by going to the total-spin basis. Moreover, because of $SU(2)$ invariance, the S matrix will be constant within every subspace of total spin, so that the information in the S matrix reduces to three scattering phases for every value of the total spin. As only the eigenstates with total spin $S=1$ contribute to the spectral function, the scattering phase in this sector is plotted in Fig.~\ref{fig:smatrix}, showing a drastic rotation of the scattering phase right at the point where the spectral function has its resonance. This confirms that the resonance is indeed a consequence of a strong attractive triplet-triplet interaction.

\begin{figure*} \centering
\includegraphics[width=1.7\columnwidth]{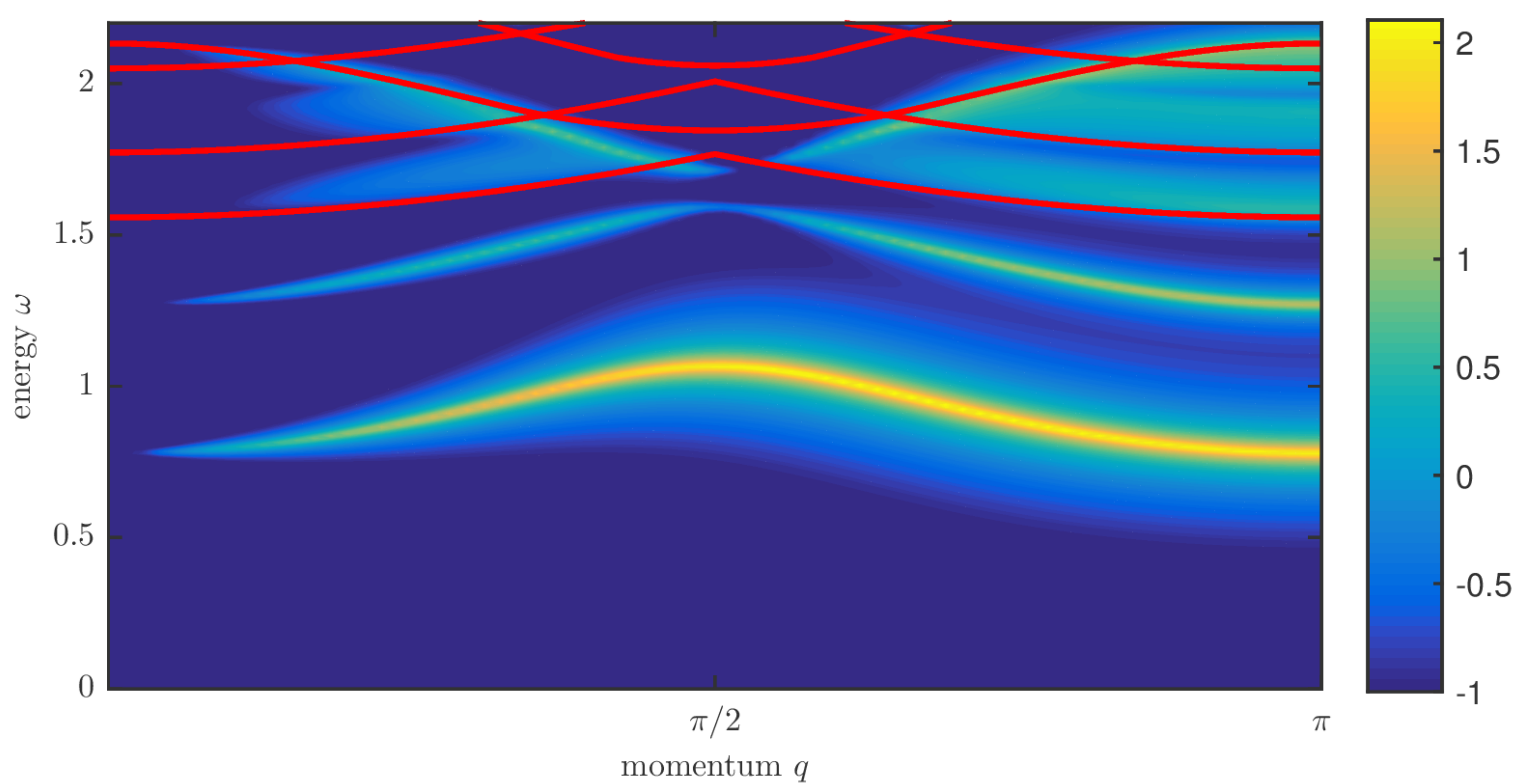}
\caption{The logarithm of the spectral function $\log_{10}S(q,\omega)$ in the full Brillouin zone for $J_1=1$, $J_2=1/2$ and $\delta=0.13$. A small imaginary frequency $\epsilon=0.01$ is added, such that the one-particle $\delta$-peaks can be plotted as Lorentzians. We have also plotted the edges of the different two-particle continua in red.}
\label{fig:spectral2}
\end{figure*}

For completeness' sake, in Fig.~\ref{fig:spectral2} we have plotted the full spectral function for $J_1=1$, $J_2=1/2$ and $\delta=0.13$, including the one-particle contributions. For these parameters only two $s\bar{s}$ bound states are stable around the minimum of the two-particle band. Around the maximum there is still a bound state (see Fig.~\ref{d}).

\section{Conclusions}

In conclusion, we have illustrated the power of a new method for computing spectral functions for generic one-dimensional quantum spin systems with unprecedented resolution in momentum and frequency. More specifically, our method is especially suited for studying quasi-particle interactions which lead to sharp resonances in the two-particle spectral function and/or the formation of bound states. The two-particle S matrix, unavailable to other methods, provides an additional tool to characterize the interactions.
\par Showing its versatility, we have applied our method to study the effects of soliton confinement on the low-lying spectral function. It was shown that the confining potential induces, in addition to the characteristic stack of stable bound states, a very specific fine structure of the spectral function in the two-particle continuum. The method allowed e.g. to reveal for the first time a transfer of spectral weight into the two-particle continuum {\it before} a bound state reaches its lower edge. We suggest that such subtle effects could be observed e.g. in INS of spin-Peierls systems by e.g. varying the external pressure in order to monitor the effective coupling $\delta$ via a change of the interchain couplings.  We also believe that similar subtle features should be seen in double-magnon Raman scattering \cite{Els1997}, a ($K=0$ singlet) four-particle response function, but this is left for future studies.

\begin{acknowledgments}
The authors would like to thank Ian Affleck for inspiring discussions. L.V. would like to thank D.P. for his hospitality in Toulouse. L.V. and J.H. are supported by the Research Foundation Flanders (FWO), F.V. by the Austrian FWF SFB grants FoQuS and ViCoM and the European grants SIQS and QUTE, and D.P. is supported by the NQPTP ANR-0406-01 grant (French Research Council).
\end{acknowledgments}

\bibliography{bibliography}

\end{document}